\documentclass[authoryear, 5p, twocolumn, 11pt]{elsarticle}
\usepackage{lipsum}
\usepackage{amsfonts, amsmath, bm, amsthm, amssymb}
\usepackage{graphicx}
\usepackage{xcolor}
\usepackage{subfigure}
\usepackage{parskip} 
\usepackage{setspace}
\usepackage{booktabs, multirow, makecell, colortbl}
\usepackage[font=scriptsize]{caption}
\usepackage{comment}
\usepackage[super]{nth}

\newcommand{\ra}[1]{\renewcommand{\arraystretch}{#1}} 

\usepackage[font=scriptsize]{caption}
\usepackage[hyphens]{url} 
\usepackage{hyperref}
\hypersetup{
    urlcolor  = black,
	colorlinks = true,
	citecolor = blue, 
	linkcolor = blue 
}

\makeatletter
\def\ps@pprintTitle{%
 \let\@oddhead\@empty
 \let\@evenhead\@empty
 \def\@oddfoot{}
 \let\@evenfoot\@oddfoot}
\makeatother

\newif\ifrev
 \revtrue
\ifrev
    
    \newcommand{\ale}[2][cyan]{{\color{#1}[ALE] #2}}
    
    \newcommand{\nic}[2][teal]{{\color{#1}[NIC] #2}}
    \newcommand{\gab}[2][orange]{{\color{#1}[GAB] #2}}
\else
    
    \newcommand{\ale}{}
    
    \newcommand{\nic}{}
    \newcommand{\gab}{}
\fi

\begin{document}

\title{Evaluative Item-Contrastive Explanations in Rankings}

\author[1]{Alessandro~Castelnovo\fnref{dr}}
\ead{alessandro.castelnovo@intesasanpaolo.com}

\author[1]{Riccardo~Crupi\fnref{dr}}
\ead{riccardo.crupi@intesasanpaolo.com}

\author[1,3]{Nicolò~Mombelli\fnref{dr}}
\ead{nicolo.mombelli@intesasanpaolo.com}

\author[1]{Gabriele~Nanino\fnref{dr}}
\ead{gariele.nanino@intesasanpaolo.com}

\author[1]{Daniele~Regoli\fnref{dr}}
\ead{daniele.regoli@intesasanpaolo.com}

\address[1]{Data Science \& Artificial Intelligence, Intesa Sanpaolo S.p.A., Italy}


\address[3]{Dept. of Economics and Management, Univ. Brescia, Italy}


\fntext[dr]{The views and opinions expressed are those of the authors and do not necessarily reflect the views of Intesa Sanpaolo, its affiliates or its employees.}

\begin{abstract}
The remarkable success of Artificial Intelligence in advancing automated decision-making is evident both in academia and industry. Within the plethora of applications, ranking systems hold significant importance in various domains. This paper advocates for the application of a specific form of Explainable AI --- namely, contrastive explanations --- as particularly well-suited for addressing ranking problems. This approach is especially potent when combined with an Evaluative AI methodology, which conscientiously evaluates both positive and negative aspects influencing a potential ranking. Therefore, the present work introduces Evaluative Item-Contrastive Explanations tailored for ranking systems and illustrates its application and characteristics through an experiment conducted on publicly available data.
\end{abstract}


\maketitle


\section{Introduction}





In today's landscape, the practice of ranking individuals has become ubiquitous and pervasive. This ranking process finds its application in a multitude of scenarios, ranging from determining creditworthiness \citep{Hicham2022}, suitability for college admissions or employment, or even assessing attractiveness in the context of dating \citep{asudeh2019designing}. Unlike traditional scenarios where the objective is to categorically differentiate, e.g., between suitable and unsuitable items, ranking involves the arrangement of items based on their relative merits. This distinction becomes particularly relevant in contexts where a constrained number of items can be accommodated within the final selection.

Consider the scenario of a traditional credit application process. Normally, financial institutions endeavor to provide loans to all clients with the capacity to repay the borrowed funds. However, the decision-making approach shifts when the bank is constrained by a predefined limit on the number $k$ of clients to whom loans can be granted. This constraint necessitates a ranking process --– the task of prioritizing potential borrowers based on their creditworthiness. This involves not only distinguishing appropriate clients from inappropriate ones but also arranging them in an order that matches the $k$ available loan slots. In essence, ranking emerges as a mechanism to optimize the allocation of resources, especially when the number of deserving individuals exceeds the allocation capacity.

It's worth noting that while the concept of ranking is closely related to recommendation, the two are not interchangeable. Recommendation systems refer to applications in which users exclusively access the top-$k$ items that result from the ranking process –- e.g., the top-5 suggested movies \citep{Vigano2023-VIGTRT, zhang2021artificial}. On the other hand, the ranking problem concerns scenarios where the interest is placed on the top-$k$ elements, but users have access to information about and the ranks of all the items under consideration. A similar distinction is implied by \cite{Competitive2022}, who talk about non-competitive vs. competitive ranking problems. This work focuses on the latter context, and as a result, our assertions cannot be directly applied to pure recommendation problems. 

As it is the case for many automated decision-making, Machine Learning (ML) currently represents one of the best alternatives to generate efficient and optimized rankings, and it has become a prevalent practice to address this challenge \citep{rahangdale2019MLmethods}. As ML models increasingly interact with humans, who hold the ultimate decision-making responsibility, the concept of eXplainable AI (XAI) has gained significant prominence alongside the advancement of AI technologies \citep{BARREDOARRIETA202082}. Indeed, ML models, in particular the most advanced ones, such as Deep Neural Networks, are opaque concerning the mechanisms through which their output is related to given inputs: this is the well-known black-box effect. XAI research represents the effort to come up with techniques to make the outcomes of non-interpretable machine learning models more and more comprehensible, facilitating the inclusion of human involvement through \textit{Human-in-the-loop} \citep{wu2022survey}, \textit{Human-on-the-loop} \citep{Ontheloop}, and \textit{Human-in-Control} \citep{HIC} approaches.  Moreover, XAI could help reduce biases resulting from the use of AI systems, allowing for an ethical analysis of the model in use \cite{arrieta2020explainable}. Particularly in the context of ranking, a common bias, known as \textit{position} bias \citep{joachims2017accurately}, emerges due to an item's position: higher-ranked items significantly influence user perception, being more likely to be examined and selected by users, even in cases of unreliable system \citep{gupta2021online, zehlike2021fairness}.

Significant advancements have emerged in the development of XAI techniques that align with human cognitive processes, such as contrastive and counterfactual explanations \citep{wachter2017counterfactual, dhurandhar2018explanations}. Specifically, contrastive explanations facilitate human comprehension by shedding light on the rationale behind choosing one outcome over another. This form of explanation is widely recognized as both effective and easily understandable. However, these explanations are well-defined and formalized primarily for classification problems, where the contrastive explanation is based on the relative position of instances with respect to the decision boundary. In the context of rankings, to the best of our knowledge, this concept still lacks proper formalization. 

\paragraph{Contributions}
The main contribution of this work is to introduce and formalize contrastive explanations in the context of ranking problems. To enhance the support for human decision-making while mitigating the impact of position bias,  we align our formalization with the paradigm of Evaluative AI, proposed by \cite{miller2023explainable}. We call such an approach \textit{evaluative item-contrastive explanation}. While we demonstrate a practical implementation using a simple linear model, we believe the steps detailed in our approach can be easily generalized and applied to black-box models.


\section{Background on Explainability} 
\label{sec:background}

Understanding the reasons for explanations, the characteristics of a good explanation, and the distinction between contrastive and counterfactual explanations provides the necessary groundwork for formalizing our proposal to enhance interpretability in ranking systems. 
Furthermore, we will present the recent line of work on explainability towards Evaluative AI, a paradigm we believe is the most suitable to follow in shaping our proposal.

\subsection{Reasons for Explanations}
\label{sec:reasons}

Explanations of outcomes in a decision-making process are useful from several perspectives, including \citep{adadi2018peeking}:
\begin{itemize}

    \item \textit{Explain to justify}: to justify the decisions made using an underlying model. Explaining the reasons behind decisions enhances their justifiability and helps build trust among stakeholders.

    \item \textit{Explain to discover}: to support the extraction of novel knowledge and the discovery of new relationships and patterns. By analyzing the explanations provided by AI systems, researchers can grasp hidden mechanisms and gain a deeper understanding of the data and underlying processes and phenomena.
    
    \item \textit{Explain to control}: to enhance the transparency of an outcome, proactively confirming or identifying potential issues. Understanding system behavior provides increased visibility over potential vulnerabilities and flaws, facilitating rapid error identification and correction. This enhanced control empowers better system management

    \item \textit{Explain to improve}: to aid scholars and practitioners in improving the accuracy and efficiency of their models. By analyzing the explanations, insights can be gained on how to enhance the model's performance and make it more effective in its intended task.
    
\end{itemize}

\subsection{What is a Good Explanation}
\label{sec:good_XAI}

We here rely on the analysis proposed by \cite{miller2019explanation}, according to which humans perceive an explanation as good when it possesses four key properties: to be \textit{contrastive}, \textit{selected}, \textit{social}, and \textit{not to rely on probabilities} and statistical relationship when presenting explanations.

\textit{Contrastive} explanations are designed to shed light on why a particular input yields a specific output \textit{instead of} an alternative output \citep{hilton1990conversational}. They provide insights into the factors differentiating the chosen outcome from alternative possibilities. 

Providing \textit{selected} explanations means that good explanations should not include \emph{all} the reasons that are causing an output: giving a complete account would typically require too much information, most of which would not be relevant for a given context and purpose. Therefore, a good explanation should consist of a limited but relevant subset of such information. People generally expect explanations that offer a concise and focused account of causative factors, as excessively lengthy explanations might be confusing or challenging to grasp. Existing work has already looked at selecting which features in the model were important for a decision, based on local explanations \cite{baehrens2010explain, robnik2008explaining} or on information gain \cite{kulesza2013too,kulesza2015principles}.


Explanations are more effective when they are set in the landscape of the recipient's existing beliefs and values. It's crucial for explanations to be tailored to the \textit{social} context of the evaluator \cite{hilton1990conversational}. This entails not only fitting the individual's knowledge but also accommodating their self-perception and surroundings. In some cases a mismatch between explanation and expectation can lead to under-reliance and significant loss of trust  despite AI system performance \cite{papenmeier2019model}.

To effectively communicate explanations, one should avoid incorporating probability and statistical arguments, as humans struggle with handling uncertainty \cite{miller2019explanation}. Probability and statistics don't provide a clear intuition to most individuals, thus they represent a poor strategy to explain anything.  The assessment of these four features of explanation poses challenges in terms of measurement \cite{doshi2017towards}.

\subsection{Contrastive Explanation} \label{contrastive}

The locus classicus for this notion is \cite{Lipton1990-LIPCE}. The basic idea of giving a contrastive explanation of $P$ is that of giving an explanation of why $P$ rather than $Q$. The relationship between $P$ and $Q$ is often referred to as a ``fact-foil structure'', where $P$ is the observed fact, and $Q$ represents the foil or a hypothetical fact that did not occur. In order to explain $P$, a contrastive explanation points to the differences in the causal histories of $P$ and $Q$ (that plausibly made the first event happen, instead of the second). It's important to note that contrastive explanations are inherently perspectival, being relative to the defined pair of fact-foil items. This perspectival nature implies that each explanation may vary, offering distinct informational content. It is debated whether the fact and the foil of a contrastive explanation should be incompatible. Intuitively, they may not always be incompatible. However, in our context of ranking problems, we may assume that facts and foils are indeed incompatible due to the constraints in ranking systems. E.g., when dealing with the positions of items in a ranked order, practical considerations often lead to the imposition of restrictions, such as the prohibition of two items occupying the same position in a rank. Consequently, admissible contrastive explanations are typically confined to those presenting incompatible items or events.

Different kinds of contrastive explanations are recognized in the literature~\citep{miller2019explanation}. The more common approach is \textit{P}-contrast, which consists in posing the question of why the item $a$ has a certain property $P$, rather than $Q$. This approach aligns with the standard fact-foil structure discussed above. Additional approaches include \textit{O}-contrast \citep{Weber2002-VANTLA}, inquiring why does the item $a$ have property $P$, while item $b$ has property $Q$, and \textit{T}-contrast \citep{MALANDRI2022103} where the question is: why does the item $a$ have property $P$ at time $t$, but property $Q$ at time $t+\delta$?

In our formalization tailored for the ranking problem, detailed in \autoref{sec:formalizing}, we introduce the \textit{Item}-contrast approach. This involves asking why item $a$ has been ranked higher than item $b$.

\subsection{Counterfactual Explanation} \label{counterfactual}\label{diff}

In the field of XAI, there seems to be an overlap in the concepts of \emph{contrastive} and \emph{counterfactual} explanations~\citep{stepin2021survey}. Counterfactual explanations are defined as a set of statements constructed to communicate what could be changed in the original profile to get a \emph{different} outcome by the decision-making process~\citep{wachter2017counterfactual}. Therefore, counterfactual explanations are normally considered contrastive by nature and give a source of valuable complementary information \citep{byrne2019counterfactuals}. However, they may not be epistemically equivalent. The distinction lies in the fact that in counterfactual explanations the focus is on the additional characteristic of representing a conditional clause  (\textit{``If X were to occur, then Y would (or might) occur''})~\citep{stepin2021survey}, thus adding a layer of causality on the contrastive statement~\citep{crupi2022counterfactual}.
Strictly speaking, the purpose of a contrastive explanation is to provide a meaningful account of the differences between the fact and the foil without explicitly providing a course of action for changing from the fact to the foil.

\subsection{The Evaluative AI Paradigm}
\label{sec:evaluativeAI}

Typical implementations of XAI techniques, including those providing contrastive explanations, result in systems that return the recommended output together with its (contrastive) explanation.
As argued in \cite{hoffman2022psychology}, this is likely to be a limited approach.
A more effective strategy for high-stakes decisions, would be to shift from recommendation-driven decision support to hypothesis-driven decision support, as proposed by~\cite{miller2023explainable}. \cite{miller2023explainable} calls his proposed paradigm ``Evaluative AI'': in short, instead of presenting reasons for a certain outcome or recommendation --- e.g., why item $a$ is preferable (or has a higher place in ranking) with respect to item $b$ --- evaluative implementation of contrastive explanation would present reasons \emph{for and against} each of the two items. 

Notice that Evaluative AI is still explainable AI, as \cite{miller2023explainable} clarifies. 
This paradigm is particularly well-suited for assessing and navigating trade-offs between different factors. In ranking problems, the score assigned to each item has meaning only relative to that of all the other items. For this reason, we believe ranking problems to be quite a natural setting for a contrastive explanation within the Evalutative AI paradigm. This approach is more effective for decision support because, as argued by \cite{miller2023explainable}, it aligns with the cognitive decision-making process that people use when making judgments and decisions, it has the potential to effectively reduce biases that affect decisions based on rankings. In particular, it could counteract the negative influence on the cognitive decision-making process of the position bias, which arises from the very nature of the ranking framework, especially under the no-ties assumption.

\subsection{Related Works on Counterfactual and Contrastive Explanation in ranking} \label{related}

In the domain of contrastive and counterfactual explanations, a significant portion of the literature in supervised ML is dedicated to explaining classification and regression models \citep[see, e.g.,][]{stepin2021survey, guidotti2022counterfactual}. Relatively less attention has been devoted to understanding and explaining the ranked outputs produced by these models \citep{anahideh2022local}. The most notable exceptions are the works by \cite{Salimiparsa2023Counterfactual} and \cite{tan2021}.

Even if there is considerable overlap between ML models designed for classification and those designed for ranking, we want to stress the fact that contrastive and counterfactual examples designed for classification models may not be directly applicable to ranking systems. In the context of counterfactuals in ranking, the challenge extends beyond explaining what needs to be changed in order to receive a different outcome, as it is crucial not only to understand the impact of changes in a single item on its ranking but also to discern how alterations in one item reverberate and influence the rankings of other items in the list. \cite{Salimiparsa2023Counterfactual} contributes to adapting existing counterfactual explanations to suit ranking purposes: unlike traditional methods, the proposed approach integrates the position of an item in the list, investigating how modifications to the item can impact its ranking. The main objective is to determine the minimum change required for an item to secure a different rank compared to other items on the list.

\cite{tan2021} builds on the literature of counterfactual explanations and applies it in the context of recommendation systems. In this way, they apply a contrastive-type explainable AI technique to something very similar to a ranking problem (even if not quite the same, as we argued in the Introduction).


To the best of our knowledge, no other relevant contribution to the field of contrastive and counterfactual explanations of ranking systems has been made in the literature besides the works of \cite{tan2021} and \cite{Salimiparsa2023Counterfactual}. Our work tries to fill the gap in the literature by focusing on contrastive explanations rather than purely counterfactual ones. Namely, we endeavor to adapt the existing approach for contrastive explanation specifically to the context of ranking, shifting the focus of the contrast from the foils to the items. 

\section{Formalizing Item-Contrastive Explanation in Ranking} 
\label{sec:formalizing}

\subsection{Setting the Stage: How Ranking Systems Work} \label{ranking}

We here introduce the typical setting of ranking problems. To illustrate our setup, we hereafter leverage the example of selecting candidates for a company job interview, without loss of generality. 

Let $D = \{d_1, \ldots, d_n\}$ denote the set of $n$ candidates supposed to be organised in a ranking list. Each individual is identified by a set of $p$ relevant attributes (e.g., skills) for performing a specific task, that we label with $X_i = (X_i^1, \ldots, X_i^p)$.

The goal is thus to assign to each candidate $d_{i}$ a score $Y_{i}$ that induces a (desired) permutation $\pi$ of the elements in $D$. By $j=\pi_{i}$ we denote the individual $i$ in $D$ that is assigned to the position $j$ in the rank, where $j = 1, 2, \ldots, n$. 

In general, a better position in the rank denotes a better \textit{utility} of the item for the user \cite{singh2018fairness, alimonda2023preserving}. In typical ranking applications, users are interested in a sub-ranking containing the best $k<n$ items of $\pi$. In our example, the company's hiring team (the user) is interested in sorting candidates based on their suitability for the vacant position and choosing the best $k$ to invite for an interview, $k$ being constrained by the time and resources the company can afford to spend in the hiring process. 

Item-contrastive explanations can assist the hiring manager in justifying and understanding why one candidate was positioned more favorably than another within a given pair of candidates. The goal is to enhance the freedom of action of the user by providing her information to eventually act on the members of the sub-ranking and change their position. 

\subsection{Evaluative Contrastive Reasoning} \label{sec:EvalContRes}

We want to focus on a specific type of contrastive explanation, namely the one trying to answer the following question:
``Why has item $a$ been ranked higher than item $b$?''. We want to stress that is different from the more traditional ``why \textit{fact} $P$ rather than \textit{foil} $Q$?'' kind of contrastive explanation. Indeed, in ranking problems, the position of item $a$ in the ranking list is determined not only by the score assigned to $a$ ($Y_a$), but also by the scores assigned to all the other items, or, better by the contrast between $Y_a$ and the other items' scores.
However, answering ``why has \textit{item} $a$ been ranked higher than \textit{item} $b$?'' using just the reasons in favor of the candidate ranked in the higher position, could enforce the bias to the user already given by the position in the rank. Aware of these risks, we are aligned with \cite{miller2023explainable} in thinking that this approach could shrink the space the user has for autonomous reflection, especially in cases of high-stakes decisions. As a consequence, her degree of control over the final outcome may be reduced. 
If we head back to the definition of contrastive explanation, we find that it encompasses the relevant differences between the fact and the foil for a given question. In our case, relevant are all those differences between $a$ and $b$ that mattered for the assignment of the respective rankings. In particular, an explanation for a given case would be of the form ``$a$ has been ranked in a better position than $b$ because $a$ exhibits certain attributes or characteristics that $b$ does not have (or has less)''. In order to adjust this form of explanation, we want to put the user in the position to perform further reasoning. To do so, we propose an evaluative contrastive explanation in ranking that mentions also the attributes in which item $b$ scored better than item $a$ but were weighted with lower importance by the ranking algorithm. The form we propose is, therefore, the following ``$a$ has been ranked in a better position than $b$ because $a$ exhibits certain attributes or characteristics that $b$ does not have (or has less), however, $b$ exhibits certain attributes or characteristics that $a$ does not have (or have less), albeit with relatively lower importance assigned by the system. 


\subsection{General Approach for Implementation and Application with Linear Model} \label{sec:GeneralApproach}

We focus on proposing guiding principles for the design and development of XAI approaches in line with our conceptualization of evaluative item-contrastive explanation, integrating the principles for a good explanation outlined in \autoref{sec:good_XAI}.

Additionally, we present an initial framework for implementing a good evaluative item-contrastive explainer, starting with a rank created from a Logistic Regression model. The choice of the Logistic Regression is motivated by its status of interpretable model, as acknowledged by the literature \citep{castelnovo2022fftree}. Therefore, we can showcase our proposed approach without the risk of complicating our exemplar analysis with technicalities specific to a given XAI or ML method.
Additionally, Logistic Regression is frequently utilized in conjunction with complex black box models to provide insights and explanations regarding their underlying decision-making processes \cite{guidotti2018survey}.

In particular, we formalise four steps to build an explanation that is \textit{selected}, \textit{contrasted}, and \textit{evaluative} for the case of Logistic Regression. 
The implementation of these properties involves the following tasks: (i) identifying the differences between contrasted items, (ii) determining the pros associated with each contrasted item, (iii) defining criteria to order the differences between contrasted items based on their importance, and (iv) establishing criteria for selecting the information to present to the user.

\paragraph{Identifying the differences between contrasted items} For a linear model, the overall discrepancy between items $a$ and $b$ can be expressed as:
\begin{equation}\label{eq:delta}
    \Delta_{a,b} = \sum_{d=1}^p \Delta_{a,b}^d = \sum_{d=1}^p \alpha_d (x^d_{a} - x^d_{b}),
\end{equation}
where $p$ is the total number of features, and $\alpha_d$ is the weight associated by the Logistic Regression to each feature $d$. 

\paragraph{Determining the pros associated with each contrasted item} The linear nature of Logistic Regression streamlines the identification of pros and cons for each item. A positive coefficient $\alpha_d$ in Logistic Regression, assigned to a specific feature, implies that higher values of that feature correspond to higher assigned scores. Conversely, a negative coefficient indicates an advantage --- in terms of score --- for the item with a lower value. As a consequence, a positive $\Delta_{a,b}^d$ implies a pro for candidate $a$ (and a con for item $b$) due to feature $d$, while a negative value has the opposite effect. 

\paragraph{Defining criteria to order the differences between contrasted items based on their importance} 
A natural criterion is to compute the importance of each contribution to the overall difference as 
\begin{equation}
    \lvert\Delta_{a,b}^d\rvert = \lvert\alpha_d (x^d_{a} - x^d_{b})\rvert.
\end{equation}
In this way, importance is contingent on both the magnitude of the weight assigned by the Logistic Regression $\lvert\alpha_d\rvert$ and the extent of difference between the raw feature values of the two items, namely $\lvert x^d_{a} - x^d_{b}\rvert$. 

\paragraph{Establishing criteria for selecting the information to present to the user} This step is due to the need to reach a balance between offering a concise and selected explanation to the users while at the same time providing sufficient information for them to make a well-informed evaluation of the items under investigation. Therefore, it is clear that the final configuration embedding this trade-off can only be context-dependent. However, technically, it is possible to offer configurable methods to facilitate this decision-making process. These methods could be configured to select the top $z$ features for each contrasted item or to pick the top feature that covers a minimum level of cumulative importance. Furthermore, a mixed method can be implemented: selecting the top feature that meets a minimum threshold of cumulative importance, and additionally, including a minimum number of features as pros for the item if they were not included based on the initial criterion.

The set of presented differences should not only be technically relevant but also \textit{social}. This involves both the format used to present the explanation and the relevance of the content with respect to the background knowledge of the intended user. 
This is a requirement that extends beyond the type of implementation used to generate the ranking algorithm and depends on the context of the application and its respective users. In line with the insights from \cite{CAMBRIA2023103111}, we suggest adopting natural language explanations and visual representation to expose the differences between contrasted items. This explanation should avoid the conventional reliance on percentages as relatively difficult to grasp if the user is not familiar with statistical principles.
Finally, in presenting the outcomes, is it important to emphasize the evaluative role of the explanation in supporting a cognitive approach to making a good decision. This includes not exacerbating the position bias already introduced by the system, particularly that involving the item ranked higher among the two under consideration.

\section{Experiment}
\label{sec:experiment}

We here present an experiment to elucidate what discussed so far. To this aim, we conduct an analysis based on the \textit{Campus Recruitment} dataset~\citep{campus-dataset}.\footnote{The dataset is available in open source at \url{https://www.kaggle.com/datasets/benroshan/factors-affecting-campus-placement}.}


The \textit{Campus Recruitment} dataset on academic and employability factors influencing placement, consists of records on job placement of 215 students from an Indian University campus. In particular, it contains information about students' education, from secondary school to post-graduate specialization. Other information about the education system and the working experience is also present.
More precisely, the dataset contains the following attributes for each student: \texttt{Gender} (65\% male students), \texttt{SSC\_P} (secondary education percentage grade, 10th grade), \texttt{SSC\_B} (Board of Education, Central/Others), \texttt{HSC\_P} (high secondary education percentage grade, 12th grade), \texttt{HSC\_B} (Board of Education, Central/Others), \texttt{HSC\_S} (specialization of high secondary education, Science/Art/Commerce), \texttt{DEGREE\_P} (undergraduate degree percentage grade), \texttt{DEGREE\_T} (field of undergraduate studies Comm\&Mgmt/Sci\&Tech/Others), \texttt{WORKEX} (previous work experience, binary), \texttt{ETEST\_P} (employability test percentage), \texttt{SPECIALIZATION} (type of MBA specialization: Marketing\&Finance/Marketing\&HR), \texttt{MBA\_P} (MBA percentage grade), \texttt{STATUS} (placement status, binary), \texttt{SALARY} (job salary, if any).

In our experiment, we use \texttt{STATUS} as binary target variable (1 placed, 0 not placed). The dataset counts 148 hired and 67 unemployed students. We refer to \cite{campus-dataset} for additional info on the data.

\begin{figure}[b!]
    \centering
    \includegraphics[width=.5\textwidth]{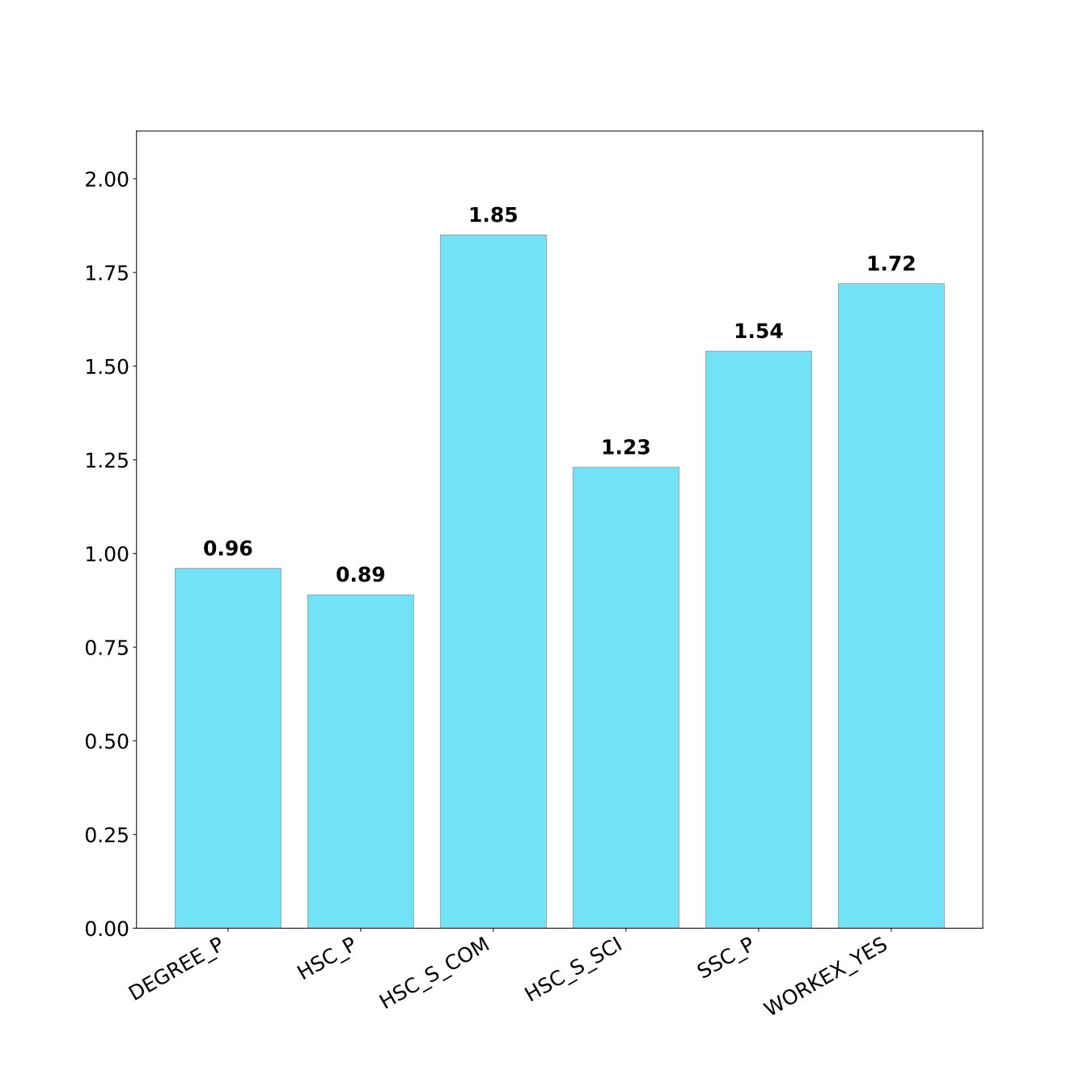}
    \caption{Coefficients of the Logistic Regression. Notice that numeric features were pre-processed via standard scaling in order to make their coefficient comparable for the evaluative phase. Moreover, categorical features have been one-hot encoded, thus e.g. \texttt{HSC\_S\_SCI} is a binary variable representing whether the student took scientific high secondary education.}
    \label{fig:coef_logit}
\end{figure}

As discussed in \autoref{sec:GeneralApproach}, we trained a Logistic Regression to predict the placement of a student given the values of the other features.
More precisely, to attain the dual goals of dropping nonsignificant features and reaching satisfactory performances, a pre-processing phase of backward step-wise feature selection has been used. In particular, we employed the $p$-value as a metric to choose the candidate feature to be removed at each step, and the significance level ($p$-value less than 5\%) as a criterion whether to retain or drop the selected feature.
\autoref{fig:coef_logit} shows the learned coefficients of the Logistic Regression with respect to the features selected via the procedure just described. Notably, the analysis reveals the significance of attending commercial or scientific programs during the high secondary school, along with the grades attained in such studies. Moreover, students with working experience seem to be strongly advantaged with the perspective of job placement. 

The model was fitted on a training set randomly sampled from the entire Campus Recruitment dataset --- 65\% of all instances, stratified with respect to the target variable in order to maintain the relative balance between the two classes.
The Logistic Regression is then applied out-of-sample on the remaining set of candidates: on the basis of the outcome, the final rank is easily determined. \autoref{tab:dataset_head} displays the rank and the significant features of the 10 candidates with the highest score. This sample will be exploited in the prosecution to provide a concrete example of item-contrastive explainability.




\begin{table*}[t!]
    \centering
    \ra{1.5}
    \caption{Top 10 candidates sorted by model's output scores (\autoref{sec:experiment}). Each entity is provided with the identification code, the ranking position and the logistic model's forecast (\texttt{SCORE}). Additionally, the most influential features as determined by the pre-processing algorithm have been included. Candidates with shaded background are those chosen to elucidate the functionality of the proposed solution as discussed in \autoref{sec:constructingEICE}.}
    \label{tab:dataset_head}
    \resizebox{.9\textwidth}{!}{
        \begin{tabular}{@{}lcccccccc@{}}
        \toprule
            \textbf{ID} & \textbf{RANK} & \textbf{SCORE} & \textbf{DEGREE\_P} & \textbf{HSC\_P} & \textbf{HSC\_S\_COM} & \textbf{HSC\_S\_SCI} & \textbf{SSC\_P} & \textbf{WORKEX\_YES} \\ 
            \midrule
            00034 & 1 & 0.99933 & 81.0 & 65.0 & 0 & 1 & 87.0 & 1 \\
            00029 & 2 & 0.99648 & 67.5 & 76.5 & 1 & 0 & 76.76 & 1 \\
            00139 & 3 & 0.9959 & 73.0 & 64.0 & 0 & 1 & 82.0 & 1 \\
            00097 & 4 & 0.99578 & 76.0 & 70.0 & 0 & 1 & 76.0 & 1 \\
            \rowcolor{black!10}\textbf{00079} & \textbf{5} & \textbf{0.99418} & \textbf{64.5} & \textbf{90.9} & \textbf{0} & \textbf{1} & \textbf{84.0} & \textbf{0} \\
            \rowcolor{black!10}\textbf{00188} & \textbf{6} & \textbf{0.9872} & \textbf{67.0} & \textbf{65.5} & \textbf{0} & \textbf{1} & \textbf{78.5} & \textbf{1} \\
            00140 & 7 & 0.98367 & 59.0 & 70.0 & 1 & 0 & 77.0 & 1 \\
            00070 & 8 & 0.98218 & 66.0 & 73.0 & 0 & 1 & 73.0 & 1 \\
            00063 & 9 & 0.9769 & 67.4 & 64.2 & 0 & 1 & 86.5 & 0 \\
            00072 & 10 & 0.97364 & 71.0 & 70.29 & 1 & 0 & 75.0 & 0 \\
            \bottomrule
        \end{tabular}
    }
\end{table*}

\subsection{Constructing Evaluative Item-Contrastive Explanations}
\label{sec:constructingEICE}


\begin{figure}[h!]
    \centering
    \includegraphics[width=.5\textwidth]{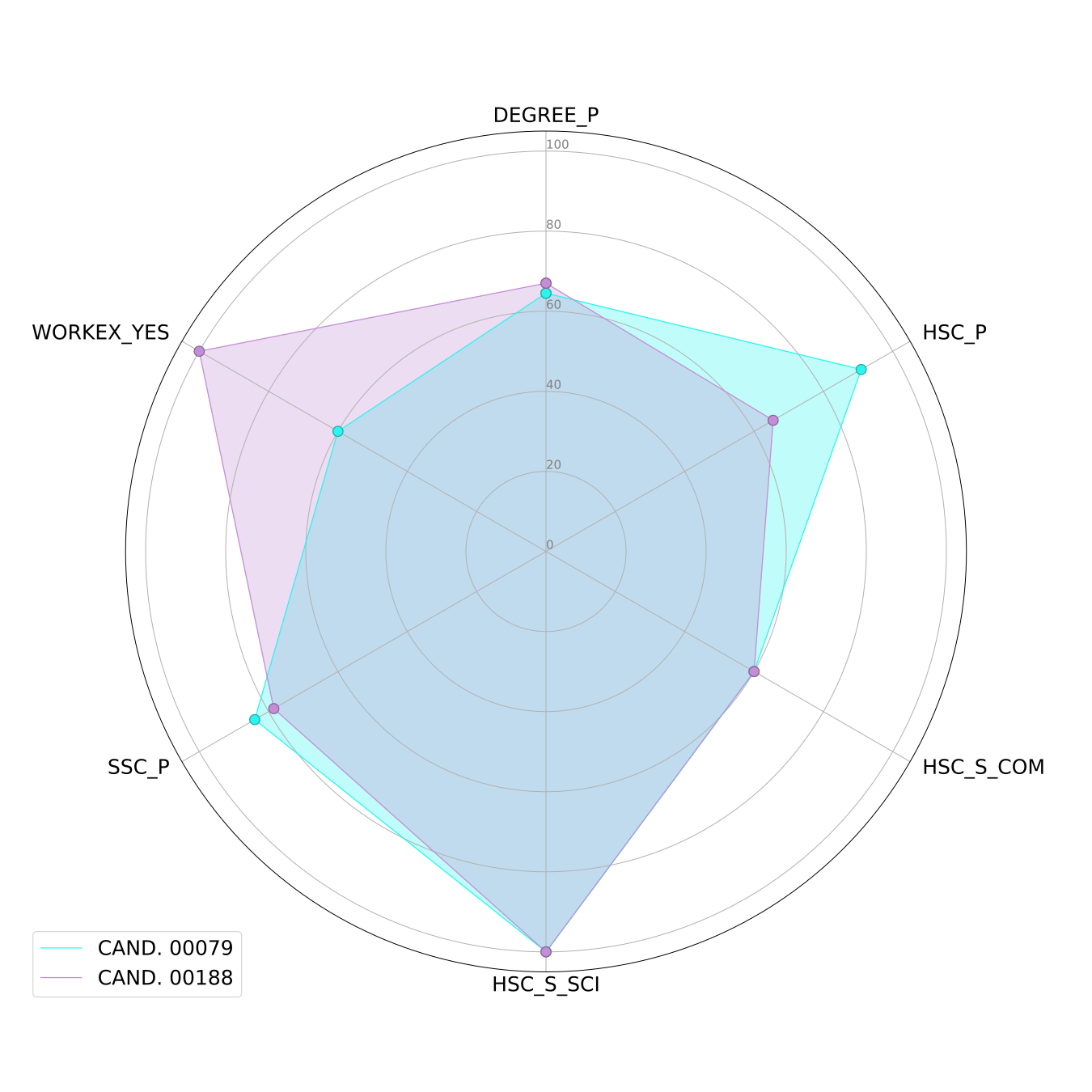}
    \caption{Juxtaposition of the key attributes pertaining to two selected candidates highlighted in \autoref{tab:dataset_head}. To enhance the clarity of feature visualization, actual feature values are presented, despite the model being trained on scaled data. Additionally, dummy variables have been encoded, with value 100 representing 1 and value 50 representing 0. An outer marker represents, for that candidate, an advantage in the hiring procedure provided by the considered feature.
    }
    \label{fig:radar_chart}
\end{figure}

\begin{figure*}[ht!]
    \centering
    \includegraphics[width=\textwidth]{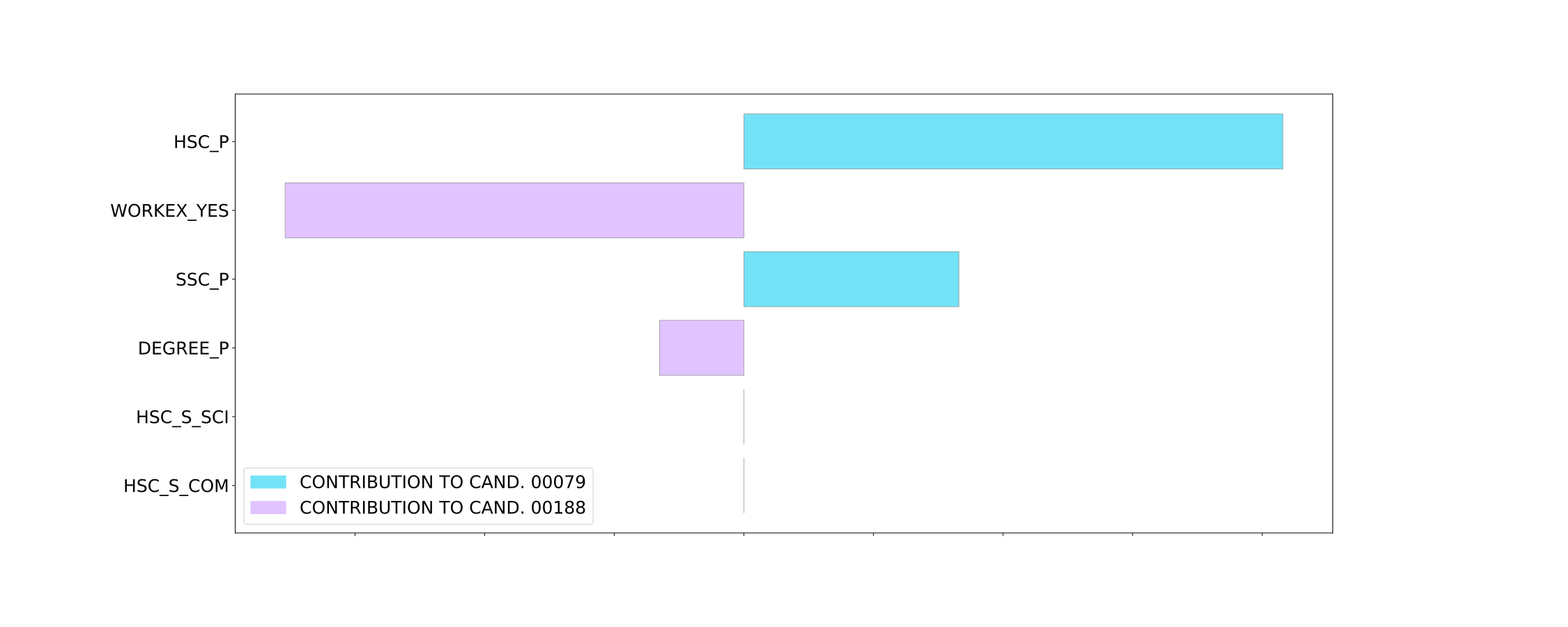}
    \caption{Contributions provided by each feature to support the disparity in ranking among two candidates highlighted in \autoref{tab:dataset_head}. Bars represent feature contributions, with length indicating the magnitude of the contribution. Bars extending to the right (left) represent contribution in favour of (in opposition to) the relative ranking. Null bars represent equal features for both candidates. Contributions are computed as percentage on the overall discrepancy.}
    \label{fig:rank_comparison_normal}
\end{figure*}

We assume that the resulting rank obtained from the Logistic Regression model represents the ordered list of candidates presented to hiring managers for selecting a limited number of $k=5$ candidates for interviews. Following our methodology, we consider an exemplar scenario in which managers start by engaging in pairwise comparisons between candidates ranked at positions 5 and 6 (see \autoref{tab:dataset_head}), and subsequently proceed with other pairs of candidates. This comparative analysis serves the dual purpose of either confirming the validity of the existing rank or potentially prompting adjustments to the final candidate selection for interviews.

In this showcase, the explanation returned by the system comprising both graphical comparisons and textual support. As mentioned in \autoref{sec:GeneralApproach}, the displayed amount of information depends on the context. In our example, given that the number of features considered by the final model has already been filtered by a feature-selection procedure, the following paragraphs outline a scenario in which all available information is provided to the user.

\autoref{fig:radar_chart} shows the comparative analysis of noteworthy attributes among candidates with ID \texttt{00079} and \texttt{00188} in \autoref{tab:dataset_head}. Notably, while both candidates pursued scientific degrees in high-secondary education, candidate \texttt{00188} concurrently gained work experience. Moreover, it is possible to retrieve that during secondary and high-secondary studies, candidate \texttt{00079} outperformed candidate \texttt{00188} academically --- based on the grades --- but the latter excelled during the bachelor's degree. \autoref{fig:rank_comparison_normal} provides a more comprehensive explanation by incorporating computed model weights, elucidating feature contributions, and their beneficiaries. Candidate \texttt{00079} predominantly benefits from having recorded higher marks during secondary education, with high-secondary education (\texttt{HSC\_P}) contributing the most and (low-)secondary grades (\texttt{SSC\_P}) approximately half of it. Conversely, candidate \texttt{00188} derives primary support from prior work experience, with a smaller contribution from higher marks in the bachelor's degree. Finally, since they both attended the same high-secondary studies (namely, scientific studies), this feature is not a discriminator among the two of them.

Alongside the visual representations, we recommend providing users with textual explanations structured as the following example:

\texttt{The available information regarding Candidate 00079 and Candidate 00188 suggests that both individuals are qualified for the job. Candidate 00079 is ranked higher than Candidate 00188 according to the current algorithm reasoning. However, the ultimate decision remains within your control, offering the option to alter this ranking if desired.
Characteristics in favour of Candidate 00079 include a higher score in HSC\_P and a higher score in SSC\_P. 
Characteristics in favour of Candidate 00188 include a higher score in DEGREE\_P and having previous working experience.}

This textual explanation is designed to empower human decision-making by promoting a comprehensive evaluation of both candidates. Its primary objective is to assist decision-makers in making informed judgments that transcend the specific information provided by the model. We advise against using judgemental (right, wrong, good, bad) or qualifying (solid, interesting, worth noting) expressions that might influence the user's perception. Moreover, expressions implying algorithmic approval could foster an overreliance on AI outcomes \citep{passi2022overreliance}. The explanation is focused solely on highlighting the differences between the items. The textual explanation, generated through an automated function, is intentionally concise, uses simple syntax, and avoids presenting numeric values and percentages.

In a real implementation, these graphical representations and textual supports could be integrated into a dashboard interface. This interactive platform allows users to dynamically engage with the model explanations, enabling them to extract essential support for informed decision-making.

\section{Discussion}
\label{sec:discussion}

As discussed throughout the manuscript, the primary objective behind our proposition of Evaluative Item-Contrastive Explanations is to support users in evaluating pairs of items, stimulating their critical judgment toward expressing a preference.

Additionally, in line with \citep{miller2023explainable}, it's essential to remember that, evaluative AI, despite its paradigm shift, is still a form of explainable AI, and as such must adhere to the standard reasons for generating explanations, as outlined in \autoref{sec:reasons}.

In this discussion, we argue that our approach remains valid also for addressing the goals of \textit{justify}, \textit{discover}, and \textit{improve}. Furthermore, within the context of ranking systems, its application becomes particularly suitable in the domains of \textit{control}.

The pros highlighted for each candidate serve as the \textit{justification} of the logic endorsed by the algorithm. Furthermore, they may contribute to the \textit{discovery} of new insights. However, attributes favoring specific items, not acknowledged by the user, could be instrumental in guiding system \textit{improvements}. When users disagree with the provided justifications, their dissent prompts a request for system enhancements.

Finally, the entire formalization of the evaluative item-contrastive explanation is crafted to empower human \textit{control} over the final decision-making process. To illustrate this, consider the following possible scenarios:

\begin{enumerate}

    \item The user agrees with the outcome justification and it is satisfied with the item's relative positions.

    \item The user agrees with the outcome justification but is unsatisfied with the item's relative positions.

    \item The user disagrees with the outcome justification but is satisfied with the item's relative positions.

    \item The user disagrees with the outcome justification and it is unsatisfied with the item's relative positions.
    
\end{enumerate}

The first case represents a scenario in which the user is in agreement with both the proposed ranking and the provided justification for why it resulted in that way. In this situation, the user will confirm the existing ordering without suggesting any improvements to the system.

In the second scenario, the user encounters a situation where she concurs with the system's provided justification but opts to modify the order of items. This circumstance, although it may initially appear counter-intuitive, underscores the importance of the evaluative paradigm, which presents the advantages and disadvantages of contrasting items. It's a clear demonstration of the user's empowerment in control: agreement with the justification indicates contentment with the system's reasoning and decision-making process. However, this doesn't imply blind adherence; the user has access to contextual elements, allowing decisions that may differ from the system's recommendations. 

The third and fourth scenarios, although resulting in different decisions from the user regarding the confirmation of the provided rank, are both marked by dissatisfaction with the justification. The user's reaction underscores their profound understanding of the organizational context and professional expertise. Their implicit knowledge justifies their preference for a specific item.  Regardless of confirmation, they recognize the need to improve the ranking algorithm, aligning its logic more closely with their well-informed judgment. However, it's essential to discuss potential directions for algorithm improvement to ensure agreement among different users, as unexpected reasoning could lead to the discovery of novel knowledge.

Summarising, considering the four potential scenarios users may encounter in ranking settings, we contend that an evaluative contrastive explanation is suitable for enhancing human oversight and control over the decision-making process.

\section{Conclusion}

In this work, we introduced and formalized the application of contrastive explanations as an effective methodology for explaining Machine Learning models for ranking. In particular, we want to stress that such approach has the merit of highlighting to the decision-maker the key elements both supporting and contrasting a proposed rank, with the ultimate goal of putting her in the most appropriate position to make an informed decision. This, in turn, helps mitigate the impact of position bias in ranking problems.

In this respect, our approach is aligned with \cite{miller2023explainable}, calling for a paradigm shift from a \emph{passive} decision-maker that can only take or reject the model's outcome, to an ever more \emph{active} decision-maker, that can truly use the model as a support to extract information on the problem at hand. 

By contrasting a pair of candidates in the proposed rank, the decision-maker can leverage granular information on what characteristics are pushing the score of one candidate above that of the other, but also what positive characteristics the lower-ranked candidate has, albeit not sufficient to be ranked higher --- given the model.

We showcased our proposal with a simple experiment on publicly available data on job placement, with an emphasis on the characteristics that explanations should possess in order to be truly effective and informative for the user. While our experiment exploits a simple linear model, thus directly using model weights as a means for extracting positive and negative contributions to the model outcomes, we plan to generalize the methodology to be used with any black-box model complemented with an arbitrary explainer providing feature contributions.

\appendix
\bibliography{ref}

\end{document}